\documentstyle[12pt]{article}
\begin{document}
\thispagestyle{empty}
\begin{flushright}
MCTP-03-05\\
%(Version of~\today)\\
\end{flushright}
\vspace{0.5cm}
\begin{center} 
{\Large \bf 
Minimal Length Uncertainty Relation and the Hydrogen Spectrum}\\[3mm]
\vspace{1.7cm}
{\sc \bf R. Akhoury and Y.- P. Yao}\\[1cm]
\begin{center} \em 
Michigan Center for Theoretical Physics\\
Randall Laboratory of Physics\\ 
University of Michigan, Ann Arbor, MI 48109-1120 
\end{center}\end{center}
\vspace{3cm}
\begin{abstract}
Modifications of Heisenberg's uncertainty relation have been proposed in the literature which imply a 
minimum position uncertainty. We study the low energy effects of the new physics responsible for this by 
examining the consequent change in the quantum mechanical commutation relations involving position and
momenta. In particular,  the modifications to the spectrum of the hydrogen atom can be naturally
interpreted as a varying (with energy) fine structure constant. From the data on the energy levels we attempt
to constrain the scale of the new physics and find that it must be close to or larger than the weak scale.
Experiments in the near future are expected to change this bound by at least an additional order of magnitude.
\end{abstract}

\vspace*{\fill}
% footnotes ------------------------------------
\noindent { e-mail: akhoury, yyao@umich.edu}
% main text ------------------------------------

\newpage
\noindent
{\bf 1.} Several independant lines of investigations appear to suggest a modification of the Heisenberg
uncertainty relation to the form:
\begin{equation}
\Delta x \geq {\hbar \over 2}\left({1 \over \Delta p} + \beta \Delta p \right),
\end{equation}
which implies a minimum position uncertainty of $\Delta x_{min}=\hbar\sqrt \beta$. In perturbative string
theory \cite{string}, such a consequence arises due to the fact that strings cannot probe distances smaller
than the string scale. We should caution however, that this particular form of the generalized uncertainty
relation  is neither a unique nor a conclusive prediction of string theory. Indeed  as shown in
ref.(\cite{kabat}), D-branes can probe distances smaller than the string scale and other generalizations of 
the uncertainty relations have been proposed, for a good review see \cite{yoneya}. Attempting to find a
general deformed Heisenberg algebra, a generalized commutation relation (in three dimensions) has been
proposed in \cite{maggiore} which implies the above uncertainty relation under certain assumptions. Finally, 
in ref.(\cite{adler}), the modified uncertainty relation has been argued to arise by taking into account the 
gravitational interactions of highly energetic photons with the other elementary particles and $\beta$
in this case is of the order of the (square of the) Planck length. In one dimension, a possible way to realize
eq.(1)
\cite{kempf,chang} is through the following commutation relation,
\begin{equation}
[X, \ \  P]=i\hbar (1+\beta  P^2)
\end{equation}
and in higher dimensions this is generalized to the following tensorial form,
\begin{equation}
[X_i, \ \  P_j]=i\hbar ((1+\beta \vec P^2)\delta_{ij}+\beta ' P_i P_j).
\end{equation}
As emphasized in \cite{chang} Eq.(1) has the interesting feature that it exhibits the so called UV/IR
correspondence first noticed \cite{uvir1} in the context of the ADS/CFT correspondence. In our case this  
is just the statement that, as is evident from eq.(1), at large $\Delta p$, $\Delta x$ is also large.
Such a correspondence is hard to visualize in the context of local quantum field theories but it seems to
be an essential feature of certain types of new physics \cite{uvir2} being considered recently.

In this paper we will take the viewpoint that eq.(1) and the commutation relation provide an effective
description by means of which new physics effects may be manifested at low energies. The fact that these
satisfy the UV/IR relation only makes this more likely and our purpose is to investigate how these kinds of
novel ideas may be manifested in particular in the well known hydrogen atom problem. A calculation of the
hydrogen spectrum indicates that in the generalized framework one is led naturally to a varying (with
energy) fine structure constant. Precision data available on the energy levels can be used to constrain
the  scale of the new physics. Writing,
\begin{equation}
\beta~=~{\bar{\beta} \over (\Lambda c)^2} 
\end{equation}
and a similar equation for $\beta'$,
with dimensionless $\bar{\beta}$ and $\bar{\beta'}$ of order one , we find that the present data implies 
that $\Lambda$ must be close to or greater than the weak scale. Within the next few years, proposed experiments
will improve this bound by two orders of magnitude. In the rest of this paper, the
details leading to the above conclusions will be elucidated. 

\bigskip

\noindent
{\bf 2.} In addition to the commutation relation (3), assuming that the momenta commute
with each other, the modified commutation relations for the coordinates are
\begin{equation}
[X_i. \ \ X_j]=i\hbar {(2\beta-\beta')+(2\beta+\beta')\beta\vec P^2
\over1+\beta \vec P^2}(P_iX_j-P_jX_i).
\end{equation}
Because the coordinates do not commute, it is more convenient to work in 
momentum space and define coordinate dependent operators through their momentum 
representations.  For the hydrogen atom, we have the Schrodinger equation
$$({\vec P^2 \over 2m}-{Ze^2\over r})|\psi>=E|\psi>,$$
or equivalently,
$$(r{\vec P^2 \over 2m}-Ze^2)|\psi>=Er|\psi>, \eqno(6)$$
which calls for a definition of r, which will be expressed as 
functions of $\vec P$, its derivatives, and the angular momentum operators 
$$L_{ij}={X_iP_j-X_jP_i\over 1+\beta \vec P^2}.$$
With this definition, one can check that the commutation relations of $L_{ij}$,
being generators of rotations are the usual ones, i.e.,

$$[X_k, \ \ L_{ij}]=i\hbar(X_i\delta _{kj}-X_j\delta _{ki}), $$

$$[P_k, \ \ L_{ij}]=i\hbar(P_i\delta _{kj}-P_j\delta _{ki}), $$
and
$$[L_{kl}, \ \ L_{ij}]=i\hbar(L_{il}\delta _{kj}+L_{ki}\delta_{lj}
-L_{jl}\delta _{ki}-L_{kj}\delta_{li}). $$

After some rather tedious algebra, one can also show that if one defines

$$\rho_i=P_if(\vec P^2),$$
where 
$$f(\vec P^2)={1\over \sqrt {(\beta +\beta')\vec P^2}}
tan^{-1}\sqrt {(\beta +\beta')\vec P^2},\eqno (7)$$
then
$$X_i=i\hbar [(1+\beta p^2){\partial \over \partial p_i}+\beta'p_i p_j 
{\partial \over \partial p_j}+\gamma p_i]
=i\hbar {\partial \over \partial \rho_i}+a(\vec p^2)p_kL_{ik}
+i\hbar \gamma p_i,\eqno (8)$$
is the representation that satisfies the relevant commutation relations 
written in eqs.(3) and (5) where, 

$$a(\vec p^2)\vec p^2=1+\beta \vec p^2-{1\over f(\vec p^2)}.$$
The constant $\gamma$ here is arbirary, which affects the 
definition of the scalar product, particularly in rendering a hermitian $X_i$. 
For a similar discussion in the context of the harmonic oscillator see \cite{chang}. Thus,

$$<f|g>=\int {d^D\vec p \over [1+(\beta +\beta')p^2]^{1-\alpha}}f^*(\vec p)
g(\vec p), \ \ \ \alpha ={\gamma-\beta'({D-1\over 2})\over (\beta +\beta')},$$
where, D is the spatial dimensionality and then,
$$<f|X_i|g>^*=<g|X_i|f>.$$

In the following, where we treat the hydrogen atom problem, we shall limit 
ourselves to D=3 and for simplicity to zero angular momentum states, which implies

$$L_i|\psi>=0, \ \ \ (g_{ij}-{p_ip_j\over p^2}){\partial\over 
\partial p_j}|\psi>=0 $$
By using eq.(8), a simple calculation leads to

$$\vec X ^2|\psi>=(i\hbar)^2[{d ^2\over d \rho^2}
+{2\over p}(1+(\beta+\gamma)p^2){d  \over d \rho}
+\gamma (3+(3\beta +\beta ' +\gamma)p^2]|\psi>,\eqno (9)$$
where
$$p^2\equiv \vec P^2, \ \ \  \rho^2=\rho_i \rho_i.$$
Our next task is to give meaning to 
$r\equiv\sqrt {\vec X^2}.$
\bigskip
We are interested in obtaining the energy eigenvalues, which 
should be independent of $\gamma$.  Therefore, to simplify the calculations 
we transform it away.  This is done by noting that

$$X_i{1\over (1+(\beta +\beta')p^2)^{\gamma\over 2(\beta+\beta')}}F=
{1\over (1+(\beta +\beta')p^2)^{\gamma\over 2(\beta+\beta')}}
X_i(\gamma=0)F.$$
We then have for eq.(9)

$$\vec X^2 |\psi>=(i\hbar)^2 ({d^2\over d\rho^2}+\phi{d\over d \rho})|\psi>,
\eqno (10) $$
where 
$$\phi={2\over p}(1+\beta p^2).$$

A change of variable is now made

$$\rho \equiv g\rho',$$
which entails

$${d\over d \rho}=
{1\over ({d \rho \over d\rho'})} {d\over d\rho'} ,\eqno (11)$$

$${d^2\over d \rho ^2}={1\over ({d \rho \over d\rho'})^2} {d^2\over d\rho'^2}
+{1\over ({d \rho \over d\rho'})} 
({d\over d\rho'}{1\over ({d \rho \over d\rho'})}){d\over d\rho'}.$$
We fix g by demanding that it satisfy the equation

$${d\over d\rho'}{1\over ({d \rho \over d\rho'})} 
 +\phi=0,$$
%Equivalently, the last equation is 

%$${d\over d \rho}(ln({d\rho' \over d\rho }))+\phi=0.\eqno (8)$$

Through this, eq.(10) becomes

$$\vec X^2 |\psi>=r^2|\psi>=(i\hbar )^2(T^{-1}{d\over d\rho'}T)
(T^{-1}{d\over d\rho'}T)T^{-1}|\psi>,$$
where 
$$T=({d\rho\over d\rho'})^2,\eqno (12)$$
The radial distance is then given by
$$r|\psi>=(i\hbar )(T^{-1}{d\over d\rho'}T)T^{-1}|\psi>.\eqno (13)$$

\smallskip
Note that because of the change of differential element implied by  eq.(7), we must likewise 
change the term without r in eq.(6) by 
$$|\psi>\to {d\rho'\over d\rho}|\psi>$$
which now becomes 

$$[i\hbar T^{-1}{d\over d\rho'}({p^2\over 2m}-E)-Ze^2({d\rho'\over d\rho})
]|\psi>=0,$$
or

$$[i\hbar {d\over d\rho}({p^2\over 2m}-E)-Ze^2
]|\psi>=0. \eqno (14)$$

For the case when we have just the conventional commutation 
relation, (i. e. $\beta=\beta'=0$, which gives $\rho =p$ 
in eq.(7)) the solutions to eq.(14) 
will immediately give the hydrogen-like spectrum, if we demand 
single-valueness of the wave functions. In fact, a more cogent 
argument is to demand that $|\psi>$ should be an entire function 
in the lower half of the complex p-plane, which will guarantee outgoing 
scattered waves and also exponentially decreasing bound state
wave functions for large r.  The derivation to  
obtain eq.(14) in that case can be made  much simpler, because when 
coordinates commute, we can use them as a representation
basis.  The representation of r in momentum space is then 
readily obtained through the Fourier representation

$$<r|p>\sim exp (ipr/\hbar),$$
which gives $r\to i\hbar{\partial \over\partial p}$ through 
integration by parts to impart its action on the momentum 
wave function.  For the extended commutation relations as 
we have, there is no $<r|$ basis, and we circumvent the 
problem through the procedure just proposed above.
\bigskip
When $\beta$ and $\beta'$ are non-vanishing, we have

$${d\over d\rho}={dp\over d\rho}{d\over dp}=(1+(\beta+\beta')p^2){d\over dp}\ ,$$
then eq.(14) beccomes 

$${d\over dp}|\psi>+{1 \over p^2-p_0^2}[2p+i{Ze^22m\over 
\hbar (1+(\beta +\beta')p^2)}]|\psi>=0 ,$$
where 

$$E={p_0^2\over 2m}.$$
Performing parital fractions, this becomes

$${d\over dp}|\psi>+[{1+i\eta \over p-p_0}+{1-i\eta \over p+p_0}
-{\xi\over p-i/\sqrt {\beta +\beta'}}+{\xi\over p+i/\sqrt {\beta +\beta'}}]
|\psi >=0,\eqno(15)$$
in which,

$$\eta={Ze^2m\over \hbar p_0(1+(\beta+\beta')p_0^2)},\eqno(16)$$ 
and, 
$$\xi={Ze^2m \sqrt {\beta +\beta'}\over \hbar (1+(\beta+\beta')p_0^2)}.$$

\bigskip
The solution to eq.(15) is 

$$|\psi>=A(p-p_0)^{-(1+i\eta)}(p+p_0)^{-(1-i\eta)}
(p-{i\over \sqrt {\beta +\beta'}})^\xi
(p+{i\over \sqrt {\beta +\beta'}})^{-\xi},$$
in which A is a constant of integration.  Note that 
for bound states, the energies are negative and therefore,
$$p_0=i\kappa .\eqno(17)$$  
The wave function will not have a pole in the lower p-plane if 
$$ \eta =-in,\eqno(18)$$
where n is a positve integer.  This is the condition to 
solve for the bound state energies.  We should remark however, that in the present 
situation with $\xi\neq 0$, we have a cut in the lower half plane for the wave function. Its effects on 
the outgoing scattered wave require further investigation, as the meaning of the radial 
coordinate r is not that intuitive. 
\bigskip
From eqs.(16),(17) and (18) we obtain for the energy levels:
$$n={Z \alpha mc \over \kappa(1-(\beta+\beta')\kappa^2)}.$$
To leading order in the small parameters using eq.(4) we obtain, for the S-wave energy levels of the hydrogen
atom 
$$E=-{Z^2\alpha^2mc^2 \over 2n^2}\left(1 + 2(\bar{\beta}+\bar{\beta'})({m \over \Lambda})^2{Z^2 \alpha^2 \over
n^2}\right). \eqno(19)$$

This may be used to define an energy dependant effective fine structure constant, 
$$\alpha_{eff}(E)=\alpha(1 - 2(\bar{\beta}+\bar{\beta'})({m \over \Lambda})^2 ({E \over mc^2})), \eqno(20)$$
to leading order in the small parameters.
Eqs(19) and (20) are the main results of this paper. We will next explore thier consequences.

\bigskip
\noindent
{\bf 3.} The 1S-2S energy shift in hydrogen has been measured to an accuracy of $1.8$ parts in $10^{14}$
\cite{hansch}. We may therefore use this to bound the scale of new physics, $\Lambda$. From eq.(19) we get,
$${E_{2S} - E_{1S} \over E_{1S}}={3 \over 4}\left(1+{1 \over 2}(\bar{\beta}+\bar{\beta'})
({m \over \Lambda})^2 Z^2\alpha^2\right).$$
Using the above mentioned experimental result \cite{hansch}, and taking $\bar{\beta}$ and $\bar{\beta'}$
to be of order unity, we obtain the bound,
$$\Lambda \geq 50 GeV.$$ 
In the near future the 2S-1S energy shift measurement is expected to be improved by two orders of
magnitude \cite{hydrogenspec} and that will bound $\Lambda$ to be greater than about a TeV if no new effects are
observed.

Since the non-standard contribution to the 2S-1S energy shift is such a small number, it would be
interesting to study its effects on the the fine and hyperfine structure. For this one will have to solve the
Dirac equation for the hydrogen atom and include the higher angular momentum states. The bound on the scale of
the new physics coming from this measurement is expected to be a whole lot stronger. Data from higher $Z$
hydrogen like ions could also potentially provide a better bound.

In conclusion, we would like to make a few comments
on some related work.  Proposing a careful measurement
of an electron in a highly excited energy level in a Penning trap,
the authors of ref.(\cite{chang}) gave a potential lower limit of $\Lambda \approx 1 Gev/c.$
This is an exceedingly difficult  endeavour.  Also, by
taking some classical limit, the same group of authors \cite{chang1} and others used
the precision limit of the precession of the perihelion of
Mercury to yield a bound of $\hbar \sqrt \beta \sim 2.3\times10^{-68}
m$, a number way below the Planck length, which {\it a fortiori }
renders the argument for a minimal length postulate suspect. As initially raised by the
authors themselves, we may
therefore question whether their approach to the classical
limit is unique in a non-commutative space. It may be safer to adhere to a pure quantum
mechanical system for an analysis, as we have done here
for the hydrogen-like atom.

In a different setting, where the space-time non-commutativity
is characterized by
$[x_\mu, \ x_\nu]=i\theta _{\mu \nu}$, where  $\theta$
is a set of external parameters, effects on
the fine structures of the hydrogen atom were estimated to yield $|\theta _{\mu \nu}|\approx 100 Gev/c$,
\cite{jabbar} which is comparable to ours.  However, it should be pointed out that rotational invariance is
violated in their considerations.  In our analysis, space is still isotropic and the hydrogen system is
a closed one.

\smallskip

After we finished this work, we were made aware that the effects of
a minimal length on the hydrogen atom had been considered by other
authors \cite{brau}.  Their results differ from ours.  Brau assumed that
$\beta '=2 \beta$ and as one can see immediately the coordinates
commute to first order in $\beta $.  Therefore, a coordinate
representation can be defined.  He took the hydrogen atom
Hamiltonian in the coordinate representation as the fundamental
one.  For us, the Hamiltonian in the momentum representation is
basic and no relationship is assumed between $\beta $ and $\beta'$.
There is no unitary transformation which connects these two
representations, if one just focuses on our momentum representation (eq.
(8))
for $X_i$  and his for momentum operator in x-space

$$P_i \Psi(x)_{Brau}={\hbar \over i}{\partial \over  \partial x_i}
(1+\beta ({\hbar \over  i}{\vec\partial \over  \partial x})^2)
\Psi(x)_{Brau},$$
except when $\beta' =-2\beta$ also, or when $\beta=0$.
In ref.\cite{brau} shifts in energy levels are due to change in the kinectic
energy term, whereas ours are due to a change in the effective
charge, as stated earlier.  The two effects are opposite in
sign.

The choice of the momentum representation made in this paper is more appropriate for this
problem since there is a minimum position uncertainity involved. Moreover both this paper
and ref.\cite{brau} have as a basis  that the momenta are commuting and then obtain eq.(5)
from eq.(3) and the Jacobi identity. Furthur, our results are not restricted in
parameter space.

\bigskip

 We are grateful to Paul Berman and Haibin Wang for a discussion.
This work was supported by the US Department of Energy.

\end{document}